# System Level Framework for Assessing the Accuracy of Neonatal EEG Acquisition

Mark O'Sullivan, Emanuel Popovici, Andrea Bocchino,
Conor O'Mahony, Geraldine Boylan, Andriy Temko

*Abstract*— Significant research has been conducted in recent years to design low-cost alternatives to the current EEG monitoring systems used in healthcare facilities. Testing such systems on a vulnerable population such as newborns is complicated due to ethical and regulatory considerations that slow down the technical development. This paper presents and validates a method for quantifying the accuracy of neonatal EEG acquisition systems and electrode technologies via clinical data simulations that do not require neonatal participants. The proposed method uses an extensive neonatal EEG database to simulate analogue signals, which are subsequently passed through electrical models of the skin-electrode interface, which are developed using wet and dry EEG electrode designs. The signal losses in the system are quantified at each stage of the acquisition process for electrode and acquisition board losses. SNR, correlation and noise values were calculated. The results verify that low-cost EEG acquisition systems are capable of obtaining clinical grade EEG. Although dry electrodes result in a significant increase in the skin-electrode impedance, accurate EEG recordings are still achievable.

## I. Introduction

Clinical evaluation of electroencephalography (EEG) is vital across neonatal, paediatric and adult patients presenting abnormal neurological behaviour. Previous publications have shown that using EEG monitoring to aid the diagnosis of neonatal seizures drastically improves the percentage of correct seizure diagnoses compared to diagnosing seizures based on clinical signs alone [1]. However, the equipment and expertise required to conduct EEG monitoring is expensive and scarcely available. Factors such as size, cost, preparation time and interpretation skills inhibit the use of EEG monitoring in a range of cases and settings. Several attempts have been made to rectify this [2], however such devices have yet to be adopted into common clinical practice.

Developing and testing novel EEG acquisition systems for clinical purposes is difficult, as it requires placing uncertified electronics on human subjects (newborns in this case) or volunteers. This requires extensive health and safety approvals that may significantly delay the process or not be obtained [3]. Thus, the development of a method to quantitatively and accurately assess the performance of an EEG acquisition system and its components on clinical data without being dependent on human participants is required.

Research was supported in part by Wellcome Trust Seed Award (200704/04/Z/16/Z), SFI INFANT Centre (12/RC/2272) and TIDA (17/TIDA/504), HRB (KEDS-2017-020), and Austrian Research Promotion Agency (853482).
M. O'Sullivan, A. Temko and E. Popovici are with the School of Engineering, University College Cork, Ireland. M. O'Sullivan, A. Temko and G. Boylan are with Irish Centre for Fetal and Neonatal Translational Research (INFANT), UCC. A. Bocchino and C. O'Mahony are with Tyndall National Institute, UCC.
(email: mark.e.osullivan@umail.ucc.ie)

Assessing the accuracy of an EEG acquisition system is challenging, as the definition of the original EEG signal, noise and artifacts is required. Previous studies utilized sinusoid waveforms to assess acquisition accuracy. However, the performance varies based on the frequency of the input signal [4]. Using an arbitrary waveform generator (AWG) outputting EEG data provides a good measure for the systems performance on the data of interest [5] [6]. A phantom head can be used to model the effects of skull and scalp to enhance the authenticity of the simulation, at significant cost [3].

We have previously presented a novel hardware and software system for acquiring and interpreting EEG in a low-cost manner that requires minimal patient preparation and expertise [7]. In order to complement that work and provide a realistic model source of EEG signals, this paper develops an experimental platform that generates EEG potentials through a simulated skin-electrode interface, using readily-available laboratory hardware and low-cost components. To accurately model the skin-electrode interface, electrical models of several wet and dry EEG electrodes are obtained through in vivo experiments on multiple healthy adult volunteers and incorporated into the platform. The signals in the simulation process are monitored at multiple stages in order to accurately account for individual losses caused by the simulation process, electrodes and acquisition system. Lastly, the effective performance of the platform has been validated on clinically obtained neonatal EEG seizure data.

## II. Methods and Materials

### A. EEG Simulation Framework Overview

The simulation setup is shown in Fig. 1. An anonymised clinical neonatal EEG database from INFANT Research Centre was used. The signals are processed to have a 250Hz sampling rate, ±0.5V amplitude, and 0.5-100Hz bandwidth. A 50Hz notch filter is applied to minimise artifacts from the previous acquisition. The data is split into 30-second epochs, and 15 epochs are randomly selected. The digital data is sent to an Agilent 33220A AWG where it is converted to analogue. The output of AWG is scaled back to the original EEG amplitude (±100μV) using a voltage divider. The signal is passed on to conductive cloth and the EEG electrodes are placed at the opposite end. The passive electrical impedance models of the electrodes are attached to the output of the respective electrodes. The EEG acquired by each electrode is connected to a separate channel on the OpenBCI Cyton Biosensing board, streaming the data into Matlab for comparison with the original data and metric computation. Ideally, neonatal intracranial EEG would have been used, however such data is not readily available. Regardless, EEG degradation due to skull artifacts is less severe with neonates due to lowered skull impedance [8].

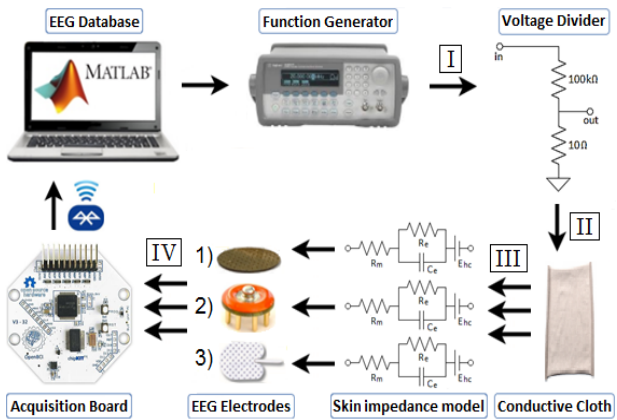

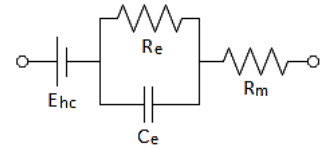

Figure 1. Simulation experiment setup.

Figure 2. Biopotential electrode equivalent circuit.

*B. Electrodes*

The current protocol used to acquire EEG requires extensive patient preparation due to the use of wet electrodes [9]. This entails the use of skin abrasives to abrade the top layer of the skin, and conductive gels to improve conductivity. Electrode-skin impedances of less than 10kΩ are achievable using this method. However, this process requires significant time and expertise. Furthermore, skin abrasion can occasionally cause irritation [10]. Novel designs have drastically simplified the electrode application process by encompassing 8 or 12 electrodes in a single structure that is pre-prepared with conductive and abrasive creams [11]. Dry electrodes require no abrasive or conductive gels, but rely on novel mechanical designs to achieve high quality contact with the skin. Preliminary studies of alternative dry electrode technologies provided promising results [12]. This study further develops these results and creates an accurate electrical model of each electrode and analyses the performance of each electrode used in a low-cost acquisition system. The following electrodes are considered in this study as shown and numbered in Fig. 1.

*1) MicroTIPs*

Micro Transdermal Interface Platforms (MicroTIPs) bypass the outermost layer of skin, which largely consists of non-conductive skin cells [13]. This is achieved by using micro-machined needle structures that penetrate the Stratum Corneum, thus making direct contact with the conductive layers of the epidermis beneath. Due to their sub-millimetre height, the electrodes do not contact pain receptors or draw blood. Unlike skin abrasion, the MicroTIPs only pierce the Stratum Corneum at specific points, leaving the majority of the layer intact, reducing the risk of infection and irritation. The MicroTIPs are fabricated using silicon wafers, which are subsequently coated in gold. The electrodes are assembled with 3M RedDot packages, which are widely used in clinics.

*2) G.Sahara Dry Electrodes*

G.Sahara electrodes, developed by G.Tec, present a pin based electrode fabricated using a gold alloy. The electrode consists of 7mm or 16mm pins, which can record high quality EEG through dense hair [14].

*3) Wet Electrodes*

Ambu Neuroline 700 and Cup electrodes were used with Ten20 conductive gel and NuPrep abrasive cream throughout this study as the conventional wet electrode setup.

*C. Skin-Electrode Impedance Modelling*

The EEG signals must propagate through the skull and skin before reaching the electrode. Therefore, monitoring and maintaining low-impedance at the skin-electrode interface is essential for accurate recordings. According to the International Federation of Clinical Neurophysiology (IFCN), contact resistance should be less than 5kΩ. However, modern differential amplifiers possess input impedances in the GΩ range, thus making high skin-electrode impedances less significant [10]. Bio-impedance tests were performed on adult volunteers in order to develop electrical models of the skin-electrode interface for each electrode. The resultant models are used in the later simulation experiments.

In order to conduct the impedance study, 5 healthy adult volunteers with varying skin and hair types participated, with approval from the Institutional Clinical Research Ethics Committee. A frequency versus impedance sweep was repeated 5 times, at the frontal region (Fp1-Fp2) and occipital region (O1-O2) on each subject for each electrode type, conducting 300 frequency vs. impedance trials in total. An Agilent E4980A LCR meter was used to measure the impedance. The LCR meter injects an AC signal through one electrode and records at the second electrode. The current of the signal is limited to 100 µA in compliance with IEC60101 for patient leakage current. The error of the meter is 0.3% [15]. This is far more accurate than the AC lead-off detection method commonly used in low-cost EEG systems, which can be in the region of 20% [16]. A 200 point frequency sweep of 20-1000Hz was implemented and the resulting resistance (R) and reactance (X) values were recorded. The obtained reactance values were negative thus denoting a capacitive reactance. The capacitance (C) was calculated as:

$$C = 1/(2\pi f |X|) \quad (1)$$

Fig. 2 shows the electrical equivalent circuit of the skin-electrode interface. $E_{hc}$ is the half-cell potential between the electrode material and the ionic solutions surrounding it. $R_m$ is the electrode material resistance, which is negligible compared to $R_e$. $R_e$ and $C_e$ represent R and C, as calculated above. At low frequencies, the circuit is dominated by the $R_m+R_e$ series combination. At high frequencies, it is dominated by the $R_m+C_e$ combination [17]. Since neonatal EEG is predominantly low frequency, for the purposes of simulating skin-electrode impedance, only $R_e$ is considered.

*D. Acquisition System*

Due to the small amplitude of EEG signals (±100uV) and the variability of the signal, low-noise and high resolution electronics are required. Custom system-on-chips (SoC) have been developed specifically for EEG recording to deal with these challenges [16]. The acquisition system used throughout this paper, OpenBCI Cyton, utilizes the ADS-1299 chip to acquire low-power (5mW/channel) and low-noise (1µVpp) EEG. IFCN set instrumentation parameters that EEG acquisition systems should obey [18].

TABLE I. IFCN EEG INSTRUMENTATION STANDARDS

|        | *Fs(Hz)* | *Res(bit)* | *Rin(MΩ)* | *CMR(dB)* | *Channels* |
|--------|----------|-----------|-----------|-----------|------------|
| IFCN   | 200      | 12        | 100       | 110       | 24         |
| openBCI| 250      | 24        | 1000      | 120       | 8/16       |

TABLE II. SKIN-ELECTRODE RESISTANCE AT 31 HZ

|         | *WF* | *WO* | *GF*  | *GO* | *MF* | *MO*  |
|---------|------|------|-------|------|------|-------|
| Res (kΩ)| 8.1  | 17.6 | 198.2 | 70.7 | 24.0 | 135.8 |

TABLE III. CORRELATION & POWER-LINE NOISE RESULTS

|              | Correlation ± 95% Conf Int. | | 50 Hz noise (uV) |
|              | *Unfiltered* | *Filtered* |                  |
|--------------|--------------|------------|------------------|
| Generator    | 0.998 ± 0.005 | 0.998 ± 0.004 | 0.32 ± 0.11 |
| Resistor     | 0.997 ± 0.015 | 0.997 ± 0.015 | 0.07 ± 0.03 |
| Cloth        | 0.997 ± 0.015 | 0.997 ± 0.015 | 0.24 ± 0.11 |
| Wet Front.   | 0.996 ± 0.016 | 0.997 ± 0.015 | 0.36 ± 0.15 |
| Wet Occip.   | 0.995 ± 0.019 | 0.997 ± 0.015 | 0.51 ± 0.21 |
| G.Tec Front. | 0.867 ± 0.555 | 0.982 ± 0.094 | 4.94 ± 2.11 |
| G.Tec Occip. | 0.978 ± 0.107 | 0.995 ± 0.019 | 1.54 ± 0.65 |
| Micro Front. | 0.990 ± 0.041 | 0.996 ± 0.015 | 0.94 ± 0.40 |
| Micro Occip. | 0.881 ± 0.511 | 0.985 ± 0.076 | 4.59 ± 1.98 |

The OpenBCI board complies with all the requirements seen in Table I, bar the minimum number of channels, which as discussed in [7], is irrelevant for the desired application. The input resistance, $R_{in}$, is of particular interest in this study as this directly affects the results for the recordings obtained through high skin-electrode impedance [10].

*E. Loss calculation*

In order to quantify the losses, the signal was recorded at stages I, II, III, IV as seen in Fig. 1. This allows us to individually account for the losses in the system due to the board and electrodes. The output of the generator (Gen) (± 0.5V) was recorded using the Tektronix MSO 3032 (Stage I). The OpenBCI board recorded the following 8 channels as referenced in Fig. 1: Stage II after the resistor divider (Res), Stage III after the conductive cloth (Clo), Stage IV for {wet frontal (WF), wet occipital (WO), G.Tec frontal (GF), G.Tec occipital (GO), MicroTIPs frontal (MF), MicroTIPs occipital (MO)}. The impedances corresponding to the electrode model of the frontal/occipital locations were used. The power-line noise was calculated by computing the FFT and computing the average power at 50Hz. The amplitudes of the signals were normalized using standard z-score normalization and the SNR and Pearson correlation coefficient were calculated before and after applying a 50Hz notch filter:

$$Z_Y = \left(\frac{Y-\mu_y}{\sigma_y}\right) \times \sigma_X + \mu_X \quad (2)$$

$$snr = 20\log\left(\frac{X}{|X-Y|}\right) \quad (3)$$

$$r = \frac{1}{N-1}\sum_{i=1}^{N}\left(\frac{X_i-\mu_x}{\sigma_x}\right)\left(\frac{Y_i-\mu_y}{\sigma_y}\right) \quad (4)$$

where *X* is the emitted signal, *Y* is the received signal, *Z* is the normalized signal, μ is mean and σ is standard deviation.

III. RESULTS & DISCUSSION

Fig. 3 shows the skin-electrode resistance and capacitance (in blue and red, respectively) vs. frequency for 3 electrodes at frontal and occipital locations. The mean and 95% confidence intervals are calculated across difference trials. Table II displays the resistance values at 31Hz, as it is a commonly used frequency for impedance detection in BCI systems [16]. These values were used in the later simulations.

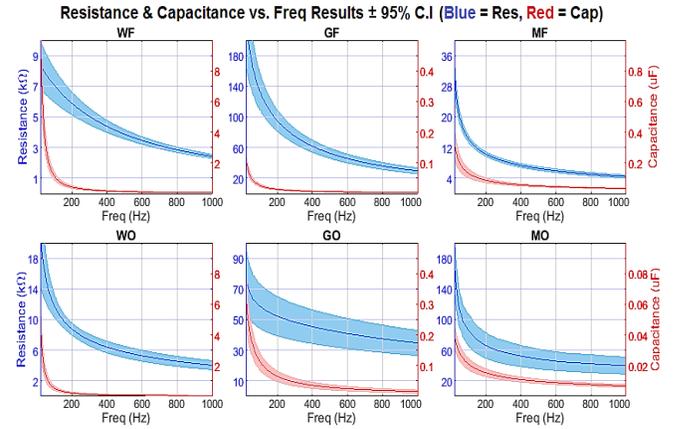

Figure 3. Skin-electrode resistance and capacitance.

As expected, the wet electrode achieves the lowest impedance due to the skin preparation via abrasive cream and improved conductivity due to conductive gel. The average impedance over to 20Hz-1kHz sweep for both frontal and occipital are below 10kΩ. The G.Sahara electrode performed better at the occipital region, over hair than on the frontal region. This is likely due to the reduced effective surface area of the conductor at the frontal region where only the 2mm wide pins make contact with the skin. At the occipital region, the conductive plate behind the pins increases the effective surface area of the conductor as conductive solutions such as oil and sweat in the hair make contact with the plate. On the contrary, the MicroTIPs achieve lower impedances on the frontal region as the Stratum Corneum (SC) is effectively penetrated, drastically reducing the impedance. These values are comparable with the gold standard wet electrodes. At the occipital region, the MicroTIPs are less effective as the 500μm tips struggle to manoeuvre through the dense hair and make contact with the SC. Thus, it is theorized that a hybrid between both dry electrodes would result in the best overall multi-location electrode.

The simulation results are presented in Table III and Fig. 4. It can be seen that the unfiltered output of the function generator achieves a near perfect recreation of the original signal, with a correlation of 0.998 and SNR of 24.9dB. Scaling the signal to μV and injecting it onto the conductive cloth introduces losses as expected, reducing the SNRs to 23.4dB and 22.7dB, respectively. The results of each electrode shows that higher impedances do increase susceptibility to noise and power-line interference, as confirmed by the power-line noise voltages.

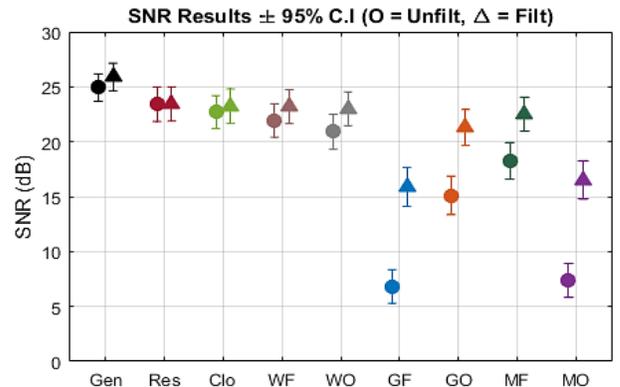

Figure 4. Simulation SNR results.

Consequently, this corrupts the EEG signal, drastically lowering the correlation to 0.867 and the SNR to 6.7dB in the case of GF, which has an impedance of 193kΩ. The increased 50Hz interference of 4.94µV in GF compared to 0.36µV for WF is largely due to the impedance mismatch reducing the common mode rejection of the differential amplifiers in the ADS1299. Applying a 50Hz notch filter to the data improves the correlation and SNR. GF achieves a correlation of 0.982 and an SNR of 15.8dB after filtering.

The obtained correlation and power-line noise amplitude compare well with previous studies, which presented correlation values of ~0.99 and power-line noise of ~0.5µV for a dry electrode with 2kΩ impedance [19]. Previous works using the MicroTIPs in the space of electrocardiography (ECG) have shown similar impedance recordings [13]. The SNR of the MicroTIPs was estimated on ECG data (±0.4 V). Values in the range of 30.3dB were obtained. These results correlate well with the observations in this study.

There are certain limitations and assumptions adopted in the experiments in this study. As previously stated, it is assumed that the artifacts introduced by the skin-electrode interface are largely due to resistive impedance. Although $R_m$ in Fig. 2 is considered negligible compared to $R_e$, previous studies have shown that the composition of the electrode material and its polarization can directly affect the signals susceptability to motion artifacts. Non-polarizable materials result in lower skin-electrode impedance at low frequencies [17]. This increased impedance at low frequencies is clearly evident in this study, as seen in Fig. 3. The impedance models used in the simulations are from adults. There are differences between adult and neonatal skin impedance, however the literature states that standard practise for both populations is to maintain impedance below 5kΩ [18] [20].

## IV. CONCLUSION

This paper describes an accurate and low-cost experimental platform intended for use in computing the accuracy of an EEG recording system and its components. The losses in the system due to the digital-to-analog conversion and down-scaling are minimal. The low-cost portable EEG acquisition board achieves high accuracy with respect to the original EEG signal. The use of dry EEG electrodes results in higher skin-electrode impedances as expected. However, micro-machined structures, such as MicroTIPs, effectively reduce the impedance, particularly in regions without hair. The use of larger pins, as with G.Sahara, reduces the impedance over the hair. Introducing these impedances into the simulation experiment has a drastic impact on the EEG signal due to impedance mismatch and signal attenuation. However, this study shows that with the use of additional filtering, large impedances do not corrupt the EEG signals enough to significantly affect the signal.

The developed evaluation framework makes use of an extensive neonatal seizure EEG database and allows us to optimize and validate novel neonatal EEG monitoring systems, such as in [21] for clinical use.

## ACKNOWLEDGMENT

The authors would like to thank G.Tec, OpenVivo and Incereb for supporting parts of this research.